\documentclass[
reprint,
superscriptaddress,
showkeys,
showpacs,
amsmath,amssymb,
aps,
prl,
]{revtex4-2}

\usepackage{braket}
\usepackage{graphicx}
\usepackage{sidecap}
\usepackage{dcolumn}
\usepackage{upgreek}
\usepackage{mathrsfs}
\usepackage{amsmath}
\usepackage{setspace}
\usepackage{bm}
\usepackage{gensymb}
\newcommand{\bigcdot}{\mathbin{\vcenter{\hbox{\scalebox{1.6}{$\cdot$}}}}}

\usepackage{tikz, mathtools}

\tikzset{dot/.style={draw, thin, circle, fill, outer sep=0pt, inner sep=0pt, minimum width=1mm}}

\usepackage[breaklinks=true,colorlinks=true,linkcolor=blue,urlcolor=blue,citecolor=blue]{hyperref}
\allowdisplaybreaks[4]
\begin{document}
	\preprint{APS/123-QED}
	\title{Boundary Time Crystals Induced by Local Dissipation and Long-Range Interactions}
	
	\author{Zhuqing Wang}
	\thanks{These authors contributed equally to this work}
	\affiliation{State Key Laboratory of Low Dimensional Quantum Physics, Department of Physics, Tsinghua University, Beijing 100084, China}
	
	\author{Ruochen Gao}
	\thanks{These authors contributed equally to this work}
	\affiliation{State Key Laboratory of Low Dimensional Quantum Physics, Department of Physics, Tsinghua University, Beijing 100084, China}
	
	\author{Xiaoling Wu}
	\affiliation{State Key Laboratory of Low Dimensional Quantum Physics, Department of Physics, Tsinghua University, Beijing 100084, China}
	
	\author{Berislav Bu\v{c}a}
	\affiliation{Universit\'{e} Paris-Saclay, CNRS, LPTMS, 91405, Orsay, France}
\affiliation{Niels Bohr International Academy, Niels Bohr Institute, University of Copenhagen, DK-2100 Copenhagen, Denmark}
\affiliation{Clarendon Laboratory, University of Oxford, Parks Road, Oxford OX1 3PU, United Kingdom}

	\author{Klaus M{\o}lmer}
	\affiliation{Niels Bohr Institute, University of Copenhagen, DK-2100 Copenhagen, Denmark}
	
	\author{Li You}
	\email{lyou@mail.tsinghua.edu.cn}
	\affiliation{State Key Laboratory of Low Dimensional Quantum Physics, Department of Physics, Tsinghua University, Beijing 100084, China}
	\affiliation{Beijing Academy of Quantum Information Sciences, Beijing 100193, China}
	
	\author{Fan Yang}
	\email{fanyangphys@gmail.com}
	\affiliation{Institute for Theoretical Physics, University of Innsbruck, Innsbruck 6020, Austria}
    \affiliation{Institute for Quantum Optics and Quantum Information of the Austrian Academy of Sciences, Innsbruck 6020, Austria}
	
	\begin{abstract}
		Driven-dissipative many-body system supports nontrivial quantum phases absent in equilibrium. As a prominent example, the interplay between coherent driving and collective dissipation can lead to a dynamical quantum phase that spontaneously breaks time-translation symmetry. This so-called boundary time crystal (BTC) is fragile in the presence of local dissipation, which can easily relax the system to a stationary state. In this work, we demonstrate a robust BTC that is intrinsically induced by local dissipation. We provide extensive numerical evidences to support existence of the BTC and study its behaviors in different regimes. In particular, with decreasing interaction range, we identify a transition from classical limit cycles to quantum BTCs featuring sizable spatial correlations. Our studies significantly broaden the scope of nonequilibrium phases and shed new light on experimental search for dynamical quantum matter.
	\end{abstract}
	
	\maketitle
	
	{\it Introduction}.---Spontaneous symmetry breaking in the temporal dimension manifests itself in a nontrivial quantum phase: the continuous time crystal (CTC) \cite{firstCTC}. In the CTC phase, a small perturbation brings the system away from the stationary state and triggers a persistent, ordered oscillation of local observables \cite{theoryTC1,theoryTC2,theoryTC3,theoryTC4,kozin2019quantum}. While existence of this intriguing phase in equilibrium has been questioned by a series of studies \cite{PhysRevLett.110.118901,doubtTC1,doubtTC2}, it has been predicted that CTC can be realized in a driven-dissipative setup \cite{ddTC1,ddTC2,ddTC4,ddTC5,ddTC6,ddTC7,alaeian2022exact,bakker2022driven,li2024time,solanki2024exotic}. This possibility is highlighted by a number of recent experiments observing self-sustained oscillations, such as in ultracold atoms \cite{coldTC1,coldTC2,dreon2022self}, thermal Rydberg gases \cite{rydbergTC1,ournp,usenp1,arumugam2025stark,jiao2024observation,wang2025time}, and solid-state materials \cite{chen2023realization,greilich2024robust}.
	
\begin{figure}[b]
		\centering
		\includegraphics[width=\linewidth]{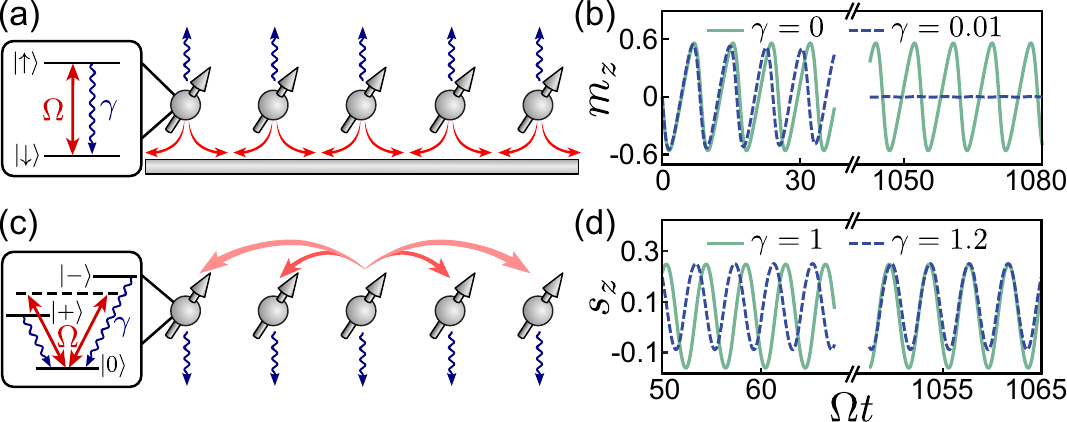}
		\caption{(a) and (b) show the schematic of the superradiance-based BTC and its mean-field evolution of the magnetization $m_z$, respectively. Here, a small local decay $\gamma=0.01\kappa$ (with $\kappa$ the collective decay rate and $\Omega=1.5\kappa$) can destroy the oscillation. (c) and (d) show the proposed spin-1 BTC model and its mean-field evolution of the order parameter $s_z$, respectively. Here, the oscillation is induced solely by local dissipation and persists within a finite range of the decay rate $\gamma$.}
		\label{fig:Fig.1}
	\end{figure}

	The dissipative version of the CTC is also called a boundary time crystal (BTC), considering the system as a boundary in contact with a bulk environment \cite{firstBTC}. The prototype model of the BTC considers a system of spin-1/2 particles subjected to a superradiant decay \cite{carmichael1980analytical}, which can be induced by a common Markovian bath [see Fig.~\ref{fig:Fig.1}(a)]. However, a realistic setup inevitably carries local decay channels, which can easily destroy the established oscillation \cite{shammah2018open,powerlawLindbladian,zhang2025emergent}, as exemplified in Fig.~\ref{fig:Fig.1}(b). The robustness of the BTC can be enhanced by a multi-level setting \cite{prazeres2021boundary}, but a local decay generally breaks the underlying $\mathrm{SU}(n)$ algebra and destabilizes the BTC.

    In this Letter, we investigate whether a BTC can be induced solely by local dissipation. The considered model is a one-dimensional (1D) spin chain, where long-range interacting spin-1 particles experience strictly local decay [see Fig.~\ref{fig:Fig.1}(c)]. While local dissipation usually erases quantum correlations \cite{eraseCorr}, we show that, remarkably, they can induce a BTC here [see Fig.~\ref{fig:Fig.1}(d)]. At the mean-field level, this BTC corresponds to a classical limit cycle, which originates from competition between distinct spin excitations. We perform extensive numerical simulations to investigate the quantum nature and origin of the BTC. In the case of all-to-all interactions, we show that the Liouvillian spectrum exhibit gapless and ordered features, while persistent oscillation of the two-time correlation function unambiguously reveals a spontaneous time-translation symmetry breaking. For a power-law interaction $\sim r^{-\alpha}$, our analysis reveals a critical $\alpha_c$, above which the interaction is too short-ranged to support the BTC. Interestingly, $\alpha_c$ is larger than the dimension of the system, giving rise to a quantum BTC without a classical correspondence in the regime $1<\alpha<\alpha_c$. In a quantum BTC, considerable spatial correlations can be rapidly established from an uncorrelated initial state.

	{\it Microscopic model}.---We consider a spin-1 model in a 1D lattice [see Fig.~\ref{fig:Fig.1}(c)], described by the Hamiltonian
	\begin{align}
		{H}=\frac{\Omega}{\sqrt{2}}\sum_i{s}_i^{x}-\Delta\sum_i{n}_i- E\sum_i{s}_i^z
		-\sum_{i<j}V_{ij}{n}_i{n}_j,
	\end{align}
	where the spin-1 operators are defined as $s_i^x = (|0\rangle_i\langle+|+|0\rangle_i\langle-|+\mathrm{H.c.})/\sqrt{2}$, $s_i^z=|+\rangle_i\langle+|-|-\rangle_i\langle-|$, and  $n_i=(s_i^z)^2=|+\rangle_i\langle+|+|-\rangle_i\langle-|$. In such a $\mathsf{V}$-type excitation scheme, the spin state $\ket{0}$ is coherently coupled to states $\ket{+}$ and $\ket{-}$ (energetically separated by $2E$) with a Rabi frequency $\Omega$ and a detuning $\Delta$ from the average energy level of the exited states. Meanwhile, spins in states $\ket{\pm}$ feature a power-law interaction $V_{ij} = C/|i-j|^\alpha$. In addition to coherent drivings and interactions, each spin in states $\ket{\pm}$ decays individually to the state $\ket{0}$ in a Markovian manner with a rate $\gamma$. The system density matrix $\rho$ then evolves according to a Lindblad master equation $\partial_t\rho=\mathcal{L}(\rho)$, where the Liouvillian operator acts as
	\begin{equation}
		\mathcal{L}(\bigcdot) = -i[H,\bigcdot] + \sum_{j=1}^N\sum_{\sigma=\pm}\left(L_{j}^\sigma\bigcdot L_{j}^{\sigma\dagger} -\frac{1}{2}\{L_{j}^{\sigma\dagger} L_{j}^\sigma,\bigcdot\}\right)
	\end{equation}
	with $L_{j}^\sigma = \sqrt{\gamma}|0\rangle_j\langle \sigma|$ the local jump operator at site $j$. This model has a neutral-atom implementation, where $\ket{0}$ denotes the ground state of an atom, and $\ket{\pm}$ represent two distinct Rydberg states \cite{browaeys2020many,wu2020concise,morgado2021quantum}, respectively.

	{\it Mean-field analysis}.---We first treat the system at the mean-field level, in which we neglect the correlation between different sites and assume that local operators $o_i$ have a homogeneous expectation value, i.e., $\langle o_i o_j\rangle=\langle o_i\rangle\langle o_j\rangle$ for $i\neq j$ and $\langle {o}_{i}\rangle=\langle {o}_{j}\rangle$. This treatment will generate a set of nonlinear equations \cite{SM}, whose nonlinear dynamics is solely determined by the collective interaction strength $\chi=\sum_{j}V_{ij}$. 
	
	We then perform stability analysis of the equations of motion \cite{nonlinearBook} and study the long-time behavior of the system. First, when states $\ket{\pm}$ are degenerate ($E=0$), the system only supports a stationary phase (SP) where observables rapidly decay to the steady-state values. Within the SP, by tuning the driving parameters $\Omega$ and $\Delta$, the system can undergo a first-order phase transition between a bistable phase and a trivial one. In fact, although $\ket{\pm}$ decay independently, the long-time behavior of the system at $E=0$ is equivalent to that of a spin-1/2 model \cite{SM}, which is widely used to describe the optical bistability in Rydberg atomic ensembles \cite{hot-bistable-1,marcuzzi2014universal,cold-bistable-1,hot-bistable-3,hot-bistable-2,zhang2024dynamical}. 
	
\begin{figure}
		\centering
		\includegraphics[width=\linewidth]{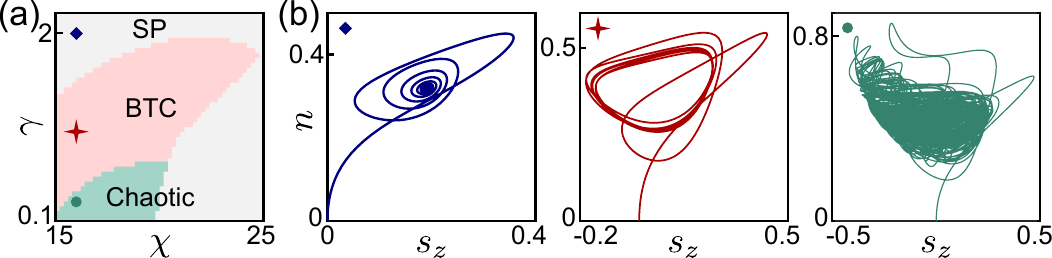}
		\caption{(a) Mean-field phase diagram in the $\chi$-$\gamma$ plane. (b) Classical trajectories of the system in the $s_z$-$n$ plane. The three panels correspond to the three data points indicated in (a). The system is initially in a product state $\rho_0=\prod_i|{0}\rangle_i\langle{0}|$.}
		\label{fig:Fig.2}
	\end{figure}

    	As the energy separation $E$ increases, the competition between states $\ket{\pm}$ comes into play, which is quantified by the order parameter $s_z=\langle s_i^z\rangle$ measuring their population difference. Near the regime ($\Omega\sim E$, $\Delta\sim-\chi/2$), the system supports a Hopf bifurcation at which the stability of a fixed point suddenly changes \cite{SM}. Figure \ref{fig:Fig.1}(d) shows a typical case $E=4$, $\Delta=-7$, $\Omega=3$, $\chi=16$, $\gamma=1$ (used throughout this work), in which $s_z$ exhibits a persistent periodic oscillation, whose trajectory is depicted by the middle panel of Fig.~\ref{fig:Fig.2}(b). The oscillation pattern of observables is akin to the coupled pendulum model \cite{hemmer1988coupled}, with the two distinct atomic coherences acting as the harmonic oscillators. This is distinct from the limit cycle in purely dissipative cases, e.g., the Lotka–Volterra predator–prey model \cite{lotka1925elements, volterra1926variazioni, jiao2024quantum}. Such an understanding is confirmed by adiabatic elimination of coherence terms: in their absence, the system ceases to oscillate.

	As we show later, the observed limit-cycle oscillation is the generic manifestation of a BTC. Such a BTC is not only robust against local decays, but also intrinsically induced by such terms, as indicated by the phase diagram in the $\chi$-$\gamma$ plane [see Fig.~\ref{fig:Fig.2}(a)]. At large decay rate $\gamma$, the system stays in the stationary phase. Here, the strong dissipation suppresses populating the interacting states $\ket{\pm}$, weakening the nonlinearity while relaxing the system to the fixed point [see the left panel of Fig.~\ref{fig:Fig.2}(b)]. As $\gamma$ becomes smaller, a robust limit cycle can be established and maintained for a finite range of $\gamma$, constituting a BTC phase. However, when $\gamma$ further decreases, the nonlinearity dominates over the dissipation and makes the system unstable, which eventually triggers a chaotic motion lacking a well-defined oscillation periodicity [see the right panel Fig.~\ref{fig:Fig.2}(b)]. Conceptually, this makes the observed BTC an intrinsic dissipative time crystal \cite{NJP}, since for a superradiance-based BTC, the ordered oscillation remains in the absence of dissipation.
	
	{\it Full quantum analysis}.---Having explored the classical limit of the model, we now turn to a full quantum description. To this end, we first assume an all-to-all ($\alpha=0$) interaction  $V_{ij}=\chi/(N-1)$, which is normalized by the system size $N$ to ensure that the interaction energy is extensive. In this regime, the system possesses a weak permutation symmetry $P_{ij}\mathcal{L}P_{ij}^{-1} = \mathcal{L}$ with respect to the permutation $P_{ij}$ between any pair of particles at sites $i$ and $j$ in a mixed state \cite{permutation, buvca2012note}. Within the symmetric subspace, the complexity of the many-body problem reduces from $\mathcal{O}(9^N)$ to $\mathcal{O}(N^8)$ \cite{SM}, allowing us to diagonalize the Liouvillian up to $N\sim 20$. While this special regime has a classical correspondence for separable initial states \cite{mattes2025long} in the thermodynamic limit ($N\rightarrow\infty$), the system is in general strongly correlated and cannot be captured by the mean-field description.
	
\begin{figure}
		\centering
		\includegraphics[width=\linewidth]{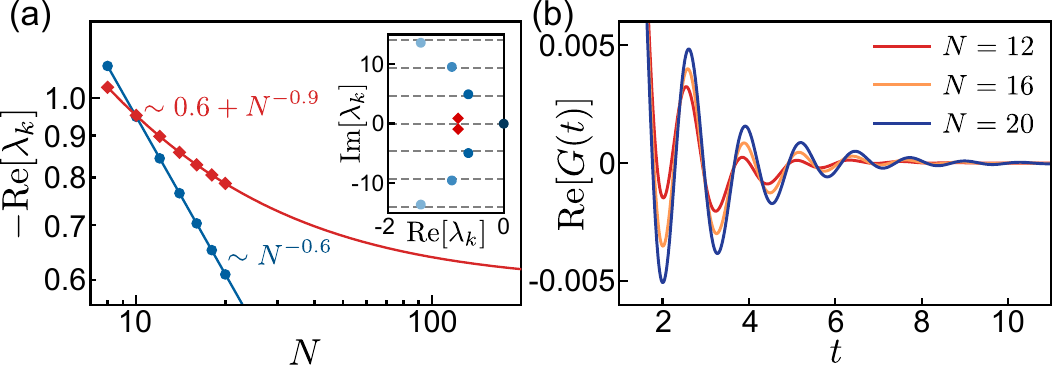}
		\caption{Characterization of the BTC with all-to-all interactions. (a) Evolution of several low-lying Liouvillian eigenvalues $\lambda_k$ with $N$ along the real axis. The data are fitted by an algebraic function $c_k+N^{-\beta_k}$. The inset shows the spectrum in the complex plane for $N=20$. (b) The two-time correlation function $G(t)$ for different system sizes.}
		\label{fig:Fig.3}
	\end{figure}

	We first study the Liouvillian spectrum $\{\lambda_k\}$ with $\mathcal{L}(\rho_k)=\lambda_k\rho_k$ in the BTC phase. As the system size increases from 10 to 20, we identify multiple ordered branches of eigenvalues $\lambda_k$ shifting towards the imaginary axis, which is a universal feature for systems supporting limit cycles \cite{dutta2025quantum}. The inset of Fig.~\ref{fig:Fig.3}(a) depicts the low-lying spectrum for $N=20$, where we can identify a branch of eigenvalues evenly spaced along the imaginary-axis direction, as indicated by the blue dots close to the dashed grids $0,\pm\omega,\pm2\omega,\cdots$, with $\omega$ the oscillation frequency of the limit cycle. This branch will gradually form a parabolic shape, responsible for a quantum-state diffusion process \cite{dutta2025quantum}, and eventually collapse onto the imaginary axis in the thermodynamic limit, generating all possible oscillation patterns of the BTC. Up to $N=20$, we find that the minimum gap of this branch follows an algebraic decay $\mathrm{Re}[\lambda]\propto -N^{-\beta}$ with $\beta\approx0.64$.
    
    In addition to the gapless branch, we find eigenvalues whose imaginary parts are incommensurate with $\omega$. The ones closest to the imaginary axis are indicated by the red diamonds in Fig.~\ref{fig:Fig.3}(a). These Liouvillian eigenstates are gapped, and determine the time scale at which a stable oscillation can be established. The gap $\approx0.59 \gamma$ drawn from an algebraic fitting shows a nice agreement with the decay rate $\approx 0.60\gamma$ toward the limit cycle.

    While the spectrum analysis shows a nice quantum-classical correspondence, we now investigate whether the phase can be defined as a BTC. For this, we focus on properties of the unique steady state $\rho_\mathrm{ss}$ for a finite $N$ satisfying $\mathcal{L}(\rho_\mathrm{ss})=0$, in analogy to the ground state of a closed many-body system. As pointed out in Ref.~\cite{doubtTC2}, a continuous time crystal can be unambiguously identified by the behavior of the two-time correlation function. For the considered open system, it is defined as 
	\begin{equation}
		G(t) = \mathrm{Tr}[\tilde{S}_z(t)\tilde{S}_z(0)\rho_\mathrm{ss}]/N^2,
		\label{eq:eq3}
	\end{equation}
	where $\tilde{S}_z=S_z-\langle S_z\rangle_\mathrm{ss}$ with $S_z=\sum_i s_i^z$ an extensive operator having a steady-state value $\langle S_z\rangle_\mathrm{ss}=\mathrm{Tr}[S_z\rho_\mathrm{ss}]$. A well-defined BTC should possess a finite and periodic correlation function $G(t)$ as $N\rightarrow\infty$ \cite{ddTC4,guo2022quantum}. Physically, $G(t)\neq 0$ implies that a finite perturbation can trigger a persistent and global oscillation, which is the key feature of a time crystal. Applying the quantum regression theorem \cite{scully1997quantum}, we obtain an increasingly persistent $G(t)$ as $N$ grows [see Fig.~\ref{fig:Fig.3}(b)]. The tail of the correlation function can be fitted into the form $\mathrm{Re}[G(t)] = Ae^{-\kappa t}\cos(\omega t+\phi)$, where $\kappa\sim N^{-0.6}$ matches the evolution of the smallest gap, while $A$ converges to a finite value, suggesting a truely persistent oscillation in the thermodynamic limit.
	
\begin{figure}[b]
		\centering
		\includegraphics[width=\linewidth]{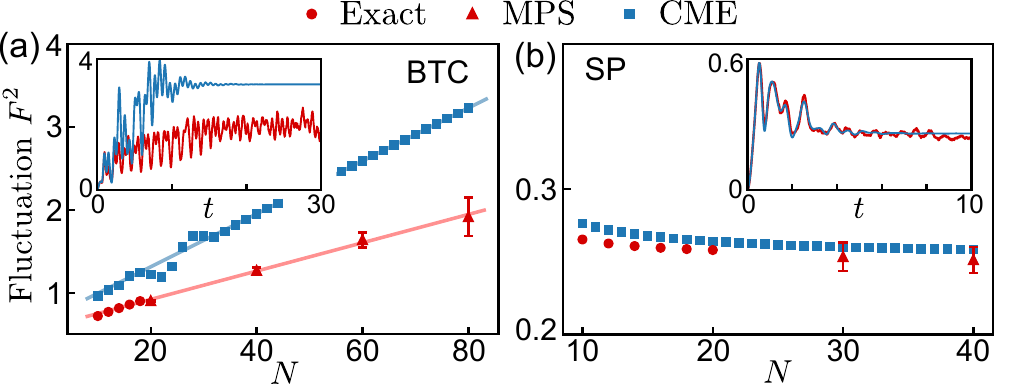}
		\caption{Scaling of the steady-state fluctuations $F^2$ with $N$. (a) and (b) show results for the BTC phase ($\Delta=-7$) and the SP ($\Delta=-1$), respectively. The insets show the time evolution of the fluctuations ($N=80$ for the BTC and $N=40$ for the SP), obtained from the MPS (red curves) and the CME (blue curves). The errorbars indicate the temporal fluctuations caused by finite trajectory realizations.}
		\label{fig:Fig.4}
	\end{figure}

	{\it Quantum fluctuations and non-Gaussianity}.---From the above calculation, we notice that the equal-time correlation $G(0)$ approaches a nonzero constant, implying that the steady-state fluctuations of $S_z$ \cite{fluctuation1,fluctuation2,fluctuation3}
	\begin{equation}
		F^2=(\langle S_z^2 \rangle_\mathrm{ss}-\langle S_z \rangle_\mathrm{ss} ^2)/N = G(0) N,
	\end{equation}
	grows linearly with the system size. Such a divergent behavior of the quantum fluctuations is a universal feature of the BTC, as was also observed in previous studies \cite{fluctuation1}. The linear scaling between $F^2$ and $N$ is verified by a larger scale simulation up to $N=80$ [see Fig.~\ref{fig:Fig.4}(a)], which is obtained by the Monte-Carlo wavefunction based matrix product state (MPS) simulation \cite{vovk2022entanglement}. For comparison, we also carry out the simulation for a stationary phase [see Fig.~\ref{fig:Fig.4}(b)]. In this regime, $F^2$ decreases with the system size and converges to a finite value, in stark contrast to the BTC phase.
	
    We next investigate the non-Gaussianity of the system by applying the cumulant expansion (CME) \cite{SM}. Up to the second order, CME assumes a Gaussian-type correlation between particles, i.e, $\langle{o}_{i}{o}_{j}{o}_{k}\rangle =  \langle{o}_{i}{o}_{j}\rangle\langle{o}_{k}\rangle + \langle{o}_{i}{o}_{k}\rangle\langle{o}_{j}\rangle + \langle{o}_{j}{o}_{k}\rangle\langle{o}_{i}\rangle -2\langle{o}_{i}\rangle\langle{o}_{j}\rangle\langle{o}_{k}\rangle$ for operators $({o}_{i}, {o}_{j}, {o}_{k})$ at three different sites. In the stationary phase, the state of the system remains largely Gaussian during the entire evolution, revealed by the good agreement between the CME and the exact/MPS simulations [see the inset of Fig.~\ref{fig:Fig.4}(b)]. However, entering the BTC phase, the discrepancies between the CME and the exact/MPS simulations become increasingly visible with a growing $N$. This implies that the evolution in the BTC phase is generally a complex problem involving intricate many-body correlations, in accordance with a recent work demonstrating the non-stabilizerness of the BTC \cite{passarelli2025nonstabilizerness}.
	
\begin{figure}
		\centering
		\includegraphics[width=\linewidth]{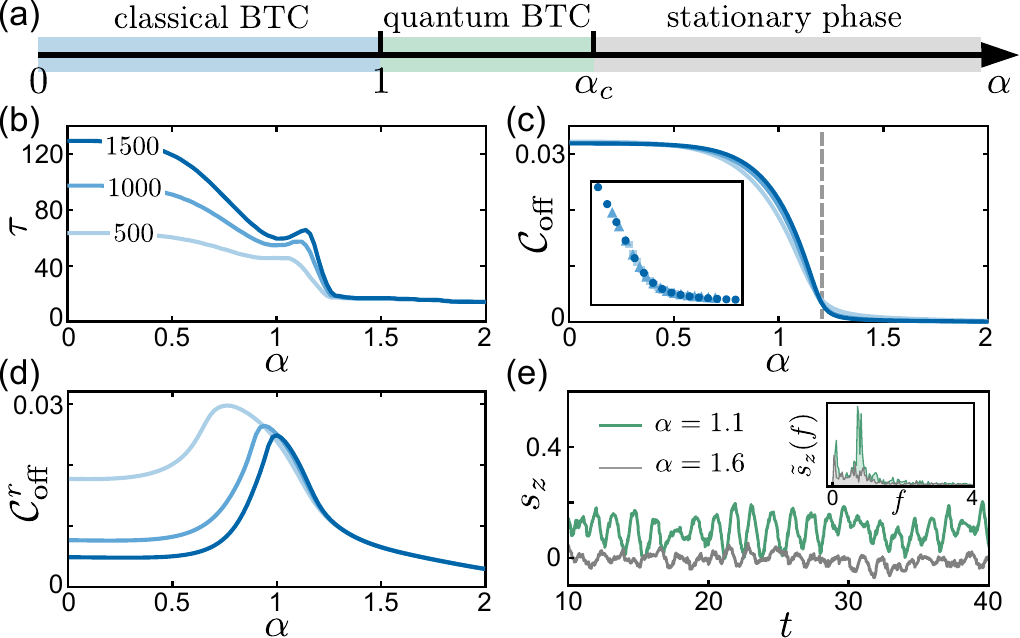}
		\caption{(a) A phase transition driven by the factor $\alpha$. (b)-(d) show the lifetime $\tau$, the steady-state off-diagonal correlation $\mathcal{C}_\mathrm{off}$, and the  mean correlation $\mathcal{C}_\mathrm{off}^{r}$ ($r=30$ and averaged over one period at $t=30$) as a function of $\alpha$ for the indicated system sizes ($N=500,1000,1500$). The results are obtained from the CME. (e) A single-trajectory MPS simulation of the dynamics for $\alpha=1.1$ (green) and $1.6$ (grey), whose Fourier spectra are shown in the inset.}
		\label{fig:Fig5}
	\end{figure}

	{\it Classical to quantum BTC}.---Despite the breakdown of exact Gaussian-type correlations, we note that the cumulant expansion can still qualitatively distinguish between the distinct scaling properties of the BTC phase and the stationary phase. We now employ this approach to investigate existence of the BTC under a power-law interaction $V_{ij}\propto |i-j|^{-\alpha}$. To ensure that the system has a constant energy per site, we apply the Kac factor to the interaction strength \cite{kac1963g}, i.e., $C=\chi/\sum_{r=1}^{N-1}{r^{-\alpha}}$. The drastically reduced complexity $\mathcal{O}(N)$ of the CME (using the translation invariance \cite{SM}) allows us to extract the phase diagram shown in Fig.~\ref{fig:Fig5}(a), which will be elucidated below.
	
	We first investigate the lifetime $\tau$ \footnote{The lifetime $\tau$ is defined as the time at which the oscillation contrast reaches a given small value $\epsilon$. While precise values of $\tau$ depend on the specific choice of $\epsilon$, different choices yield the same qualitative results described in the main text, which uses $\epsilon=10^{-5}$.} of the oscillating dynamics starting from a separable state $\rho_0=\prod_i|{0}\rangle_i\langle{0}|$. Figure~\ref{fig:Fig5}(a) shows $\tau$ as a function of $\alpha$. For a small $\alpha$, as $N$ grows from $500$ to $1500$, the lifetime gets prolonged, suggesting a closing Liouvillian gap and a BTC phase, as in the $\alpha=0$ case. However, if the interaction is too short-ranged, $\tau$ converges to a finite value, indicative of a gapped, stationary phase. 
    
    The transition from the BTC to the SP occurs abruptly at an $\alpha$ between 1 and 1.5. To study the critical behavior driven by the long-range exponent $\alpha$, we then focus on the long-range off-diagonal correlation in the steady state
	\begin{equation}
		\mathcal{C}_\mathrm{off}=\frac{1}{N(N-1)}\sum_{i\neq j}\left[\langle s_i^z s_j^z\rangle_\mathrm{ss}-\langle s_i^z\rangle_\mathrm{ss} \langle s_j^z\rangle_\mathrm{ss} \right],
	\end{equation}
	In fact, a time crystal defined by Eq.~\eqref{eq:eq3} possesses not only temporal order but also long-range spatial order \cite{souza2023sufficient}, which is indeed true for all-to-all interaction since $\mathcal{C}_\mathrm{off}\approx F^2/N=G(0)$ remains finite in the large-$N$ limit. As shown in Fig.~\ref{fig:Fig5}(c), $\mathcal{C}_\mathrm{off}$ decreases monotonically with an increasing $\alpha$ and eventually vanishes when entering the SP \footnote{This trend is verified by the MPS for $N=20$ \cite{SM}}. The critical property is then analyzed through finite-size scaling by assuming a universal form of $\mathcal{C}_\mathrm{off}=N^{\zeta/\nu}\mathcal{G}[N^{1/\nu}(\alpha-\alpha_c)]$. We obtain a nice data collapse [see the inset of Fig.~\ref{fig:Fig5}(c)] and extract a critical value $\alpha_c\approx1.22$ together with two critical exponents $(\zeta,\nu)\approx (-0.56,1.40)$. We note that $\alpha_c$ is generally larger than 1 but not rigid under variation of system parameters, typical for long-range interacting systems \cite{maghrebi2017continuous}.
	
	A critical $\alpha_c$ larger than the system dimension allows for a further classification within the BTC phase. For an uncorrelated initial state, $\alpha \leq1$ yields a locally noninteracting system as $N\rightarrow\infty$ \cite{mattes2025long}, such that evolution of local observables can be predicted by the mean-field theory. In this regime, the oscillation can be interpreted as a classical limit cycle, representing a classical BTC. However, for $1<\alpha<\alpha_c$, the interaction strength $V_{ij}$ between two arbitrary particles remains finite in the thermodynamic limit, facilitating establishment of spatial correlations at finite time. The correlation subsequently prevents a mean-field description of the oscillation, which, hence can be dubbed as a quantum BTC. To demonstrate this, we show in Fig.~\ref{fig:Fig5}(d) the mean correlation $\mathcal{C}_\mathrm{off}^{r}=\sum_{d=1}^r[\langle s_i^z s_{i+d}^z\rangle-\langle s_i^z\rangle \langle s_{i+d}^z\rangle ]/r$ within a given range $r$ during the oscillation. As $N$ increases, $\mathcal{C}_\mathrm{off}^{r}$ becomes vanishingly small for $0\leq\alpha \leq 1$, but is still nonzero for $\alpha>1$. We remark that the notion of classical and quantum BTCs is based on a dynamical classification, i.e., if the initial state $\rho_0$ carries correlations (cannot be factorized), a BTC with $\alpha\leq 1$ does not have a classical correspondence, either.
	
    The existence of a quantum BTC is verified by a single-trajectory MPS simulation, which is shown to be capable of unveiling a BTC phase for large system size \cite{cabot2023quantum}. Figure~\ref{fig:Fig5}(e) displays the MPS simulations with power-law interactions \cite{mpsevolution} for $N=100$. In the predicted stationary phase ($\alpha=1.6$), we observe a disordered oscillation of $s_z$, mainly caused by the random quantum jumps. In contrast, the oscillation in the quantum BTC regime ($\alpha=1.1$) has an obviously larger contrast, as well as a more ordered structure supported by a much more pronounced peak in the Fourier spectrum (see the inset).

{\it Conclusions and outlook}.---To summarize, we investigate the dynamical phase of an open many-body system composed of long-range interacting spin-1 particles. We demonstrate existence of the BTC in a wide parameter range, supported by an extensive analysis involving mean-field dynamics, Liouvillian spectrum, as well as temporal and spatial correlation functions. Such a highly nontrivial phase survives in the presence of local decay, making it fundamentally more robust than previously discovered BTCs. Moreover, we show that by decreasing the range of the interaction, a BTC that has a classical correspondence undergoes a transition into a quantum version, which carries nonvanishing correlations and cannot be interpreted as a limit cycle.

The proposed model can be potentially realized in artificial quantum systems, such as trapped ions or neutral atoms. In the former case, tunable power-law interactions with $0\leq\alpha<3$ can be achieved \cite{leibfried2003quantum, blatt2012quantum, monroe2021programmable}, aligning with the range considered in this work. For the latter, all-to-all interactions can be induced by cavity QED \cite{sorensen2002entangling,PRXQuantum.3.020308} or Rydberg dressing in atomic ensembles \cite{henkel2010three}. Our model can also be generalized to higher dimensions, where the long-time dynamics may be accessed by improved Krylov methods \cite{loizeau2025opening}. In higher dimensions, dipolar ($\alpha=3$) or van der Waals ($\alpha=6$) interactions offered by Rydberg-atom arrays \cite{de2019observation, semeghini2021probing, kim2018detailed} may be sufficiently long-ranged to stabilize the BTC. In addition, our studies may reveal a new universality class for nonequilibrium phases \cite{sieberer2023universality}, and open the door to exploring versatile dynamical phases, such as quasi-time crystals \cite{solanki2024chaos}, multistability \cite{he2022superradiance, bi2024folding, ma2024folded}, and measurement-induced phase transition \cite{delmonte2024measurement}.

	\begin{acknowledgments}
		{\it Acknowledgments}.---We thank T. Pohl, F. Russo, Z.C. Zhu, X.L. Li, C. Liang, Y. Zhang, G.M. Zhang for providing helpful suggestions, and X.Y. Wen, Z. Fu, C.P. Qin for sharing computation resources. F.Y. acknowledge valuable discussions with T. Vovk, L. Sieberer, and H. Pichler. This work is supported by the Innovation Program for Quantum
Science and Technology (2021ZD0302100) and by
the NSFC (Grants No. 12361131576, No. 92265205). B.B. acknowledges the support from funding by the French National Research Agency (ANR) under project ANR-24-CPJ1-0150-01 and a research grant (42085) from VILLUM FONDEN. K.M. acknowledges the support from the Carlsberg Semper Ardens project QCooL. Work in Innsbruck is supported by the ERC Starting grant QARA (Grant No. 101041435).
		\vspace{16pt}
		
		{\it Note added}: During preparation of this manuscript, we became aware of a related work on three-level dissipative time crystals induced by a dipolar interaction in 2D \cite{Thomas}.
	\end{acknowledgments}

	\bibliography{ref}

	
\end{document}


\preprint{APS/123-QED}
	\title{Supplementary Material for ``Boundary Time Crystals Induced by Local Dissipation and Long-Range Interactions''}
	
		\author{Zhuqing Wang}
	\thanks{These authors contributed equally to this work}
	
	\author{Ruochen Gao}
	\thanks{These authors contributed equally to this work}
	
	\author{Xiaoling Wu}
	
	\author{Berislav Bu\v{c}a}

	\author{Klaus M{\o}lmer}
	
	\author{Li You}
	
	\author{Fan Yang}
	
	
	
	
	
    
	\begin{abstract}
         This Supplementary material provides details of the main text, including: (i) derivation of the mean-field equations; (ii) application of the generalized permutation symmetry; (iii) cumulant expansions; (iv) quantum-trajectory-based matrix product state simulations.
	\end{abstract}

	\maketitle
	\onecolumngrid
	\noindent

	\section{Mean-field analysis}\label{AppA}
	
    In this section, we derive the mean-field equations of the system and show the corresponding phase diagram with typical parameters. The model system considered in the main text is described by the Hamiltonian
	\begin{align}
		{H}=\frac{\Omega}{\sqrt{2}}\sum_i{s}_i^{x}-\Delta\sum_i{n}_i- E\sum_i{s}_i^z
		-\sum_{i<j}V_{ij}{n}_i{n}_j.
	\end{align}
    The spin-1 operators are defined as $s_i^z=|+\rangle_i\langle+|-|-\rangle_i\langle-|$, $n_i=(s_i^z)^2=|+\rangle_i\langle+|+|-\rangle_i\langle-|$, $s_i^x = (|0\rangle_i\langle+|+|0\rangle_i\langle-|+\mathrm{H.c.})/\sqrt{2}$, $s_i^y = (|0\rangle_i\langle+|-|0\rangle_i\langle-|-\mathrm{H.c.})/\sqrt{2}$, $s_i^{-x} = (|0\rangle_i\langle+|-|0\rangle_i\langle-|+\mathrm{H.c.})/\sqrt{2}$, $s_i^{-y} = (|0\rangle_i\langle+|+|0\rangle_i\langle-|-\mathrm{H.c.})/\sqrt{2}$, $q_i=i(|+\rangle_i\langle-|-|-\rangle_i\langle+|)$, and $d_i=|+\rangle_i\langle-|+|-\rangle_i\langle+|$. The coupling between ground state $|0\rangle$ and two excited states $|+\rangle$ , $|-\rangle$ is established by the applied laser field with a Rabi frequency $\Omega$ and a frequency detuning $\Delta$. The energy difference between the two excited states $\ket{\pm}$ is $2E$. In addition, both excited states $\ket{\pm}$ decay locally to the ground state $|0\rangle$ with a decay rate $\gamma$. The system density matrix $\rho$ is governed by a Lindblad master equation
	\begin{align}
		\partial_t\rho=\mathcal{L}(\rho) = -i[H,\rho] + \sum_{j=1}^N\sum_{\sigma=\pm}\left(L_{j}^\sigma\rho L_{j}^{\sigma\dagger} -\frac{1}{2}\{L_{j}^{\sigma\dagger} L_{j}^\sigma,\rho\}\right),
		\label{Lindbladmaster}
	\end{align}
	where $L_{j}^\sigma = \sqrt{\gamma}|0\rangle_j\langle \sigma|$ denotes the local jump operator at site $j$. In the mean-field analysis, we neglect the correlation between different sites and the inhomogeneity of spatial distribution to get a closed set of equations for the first moments, e.g., $s_z = \langle s_i^z\rangle$. Based on Eq.~\eqref{Lindbladmaster}, the equations of motion for the first moments are given by
	\begin{align}
		\dot n &=\frac{\Omega}{\sqrt{2}}{s}_{-y}-\gamma n,  \nonumber\\
		\dot s_z &=\frac{\Omega}{\sqrt{2}}{s}_{y}-\gamma s_z,  \nonumber\\
		\dot s_x &=E s_y+(\Delta+\chi n){s}_{-y}-\frac{\gamma}{2} s_x,  \nonumber\\
		\dot s_y &=-\frac{\Omega}{\sqrt{2}}{s}_{z}-E s_x + (-\Delta-\chi n){s}_{-x}-\frac{\gamma}{2} s_y,  \nonumber\\
		\dot s_{-x} &=-\frac{\Omega}{\sqrt{2}}q+E s_{-y} + (\Delta+\chi n){s}_{y}-\frac{\gamma}{2} s_{-x},  \nonumber\\
		\dot s_{-y} &=-\frac{\Omega}{\sqrt{2}}(3n+d-2)-E s_{-x} + (-\Delta-\chi n){s}_{x}-\frac{\gamma}{2} s_{-y},  \nonumber\\
		\dot q &=\frac{\Omega}{\sqrt{2}}s_{-x}+2E d-\gamma q,  \nonumber\\
		\dot d &=\frac{\Omega}{\sqrt{2}}s_{-y}-2E q-\gamma d.
		\label{BlochEquations}  
	\end{align}
Here, the nonlinear dynamics is solely determined by the collective interaction strength $\chi=\sum_{j}V_{ij}$. 

In the special case $E=0$, i.e., the two excited states $|+\rangle$ and $|-\rangle$ are degenerate, these equations simplify to
	\begin{align}
		\dot n &=\frac{\Omega}{\sqrt{2}}{s}_{-y}-\gamma n,  \label{S5} \\ 
		\dot d &=\frac{\Omega}{\sqrt{2}}{s}_{-y}-\gamma d,  \label{S6} \\
		\dot s_x &=(\Delta+\chi n){s}_{-y}-\frac{\gamma}{2} s_x,  \\
		\dot s_{-y} &=-\frac{\Omega}{\sqrt{2}}(3n+d-2)+ (-\Delta-\chi n){s}_{x}-\frac{\gamma}{2} s_{-y}.\label{S8}
	\end{align}
    In view of Eqs.~\eqref{S5} and \eqref{S6}, $d$ will approach $n$ in the long-time limit ($t\rightarrow\infty$). Then, by making the substitution $d=n$ in Eq.~\eqref{S8}, we find that the mean-field equations are formally equivalent to those of a two-level Rydberg system \cite{marcuzzi2014universal}. Such a nonlinear system only supports stable or bistable phases.
	
\begin{figure}
		\centering
		\includegraphics[width=0.96\linewidth]{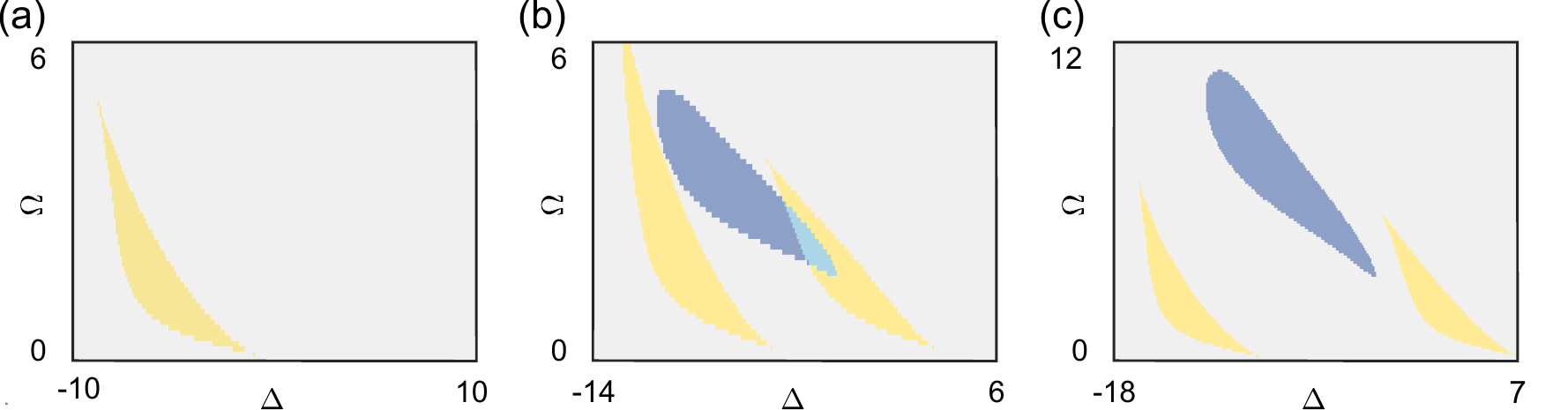}
		\caption{Mean-field phase diagram in the $\Delta-\Omega$ plane. (a) $\chi=16, \gamma=1, E=0$. (b) $\chi=16, \gamma=1, E=4$. (c) $\chi=16, \gamma=1, E=8$. The different colors in the phase diagram correspond to distinct phases, with the grey, yellow, dark blue, and light blue regions representing the SS phase, BS-1 phase (the BS-1 phase is composed of two SS phases), BTC phase, and BS-2 phase (the BS-2 phase is composed of one SS phase and one BTC phase), respectively.} 
		\label{fig:Fig.S1}
	\end{figure}

When $E\neq0$, the competition between excited states plays a role ($s_z\neq 0$), which significantly enriches the phase diagram. With a standard stability analysis, we can extract the phase diagram shown in Fig.~\ref{fig:Fig.S1}. In the degenerate case ($E=0$), as expected, only stable (SS) and bistable (BS) phases are observed [see Fig.~\ref{fig:Fig.S1}(a)]. As $E$ increases, a BTC phase emerges [see Fig.~\ref{fig:Fig.S1}(b)]. When $E$ further increases, the BTC phase gets shifted but can always be observed near the regime ($\Omega\sim E$, $\Delta\sim-\chi/2$) [see Fig.~\ref{fig:Fig.S1}(c)].
	
	\section{Exact solutions based on permutation symmetry}\label{AppB}
	For an all-to-all interaction, the system possesses a weak permutation symmetry $P_{ij}\mathcal{L}P_{ij}^{-1} = \mathcal{L}$ with respect to the permutation $P_{ij}$ between any pair of particles at sites $i$ and $j$ in a mixed state. Applying this symmetry, we can significantly reduce the computational complexity of the master equation. First of all, we can map the state transition operators $|\alpha \rangle \langle \beta|$ to symmetric operators \cite{permutation}
	\begin{equation}
		\hat{\mathbf{n}}=(n_{00},n_{0+},n_{0-},n_{++},n_{+-},n_{--},n_{+0},n_{-0},n_{-+})
	\end{equation}
	with nine positive integers in the range $[0,N]$. In general, these integers can be determined by the following formula $n_{cd}=\sum_{j=1}^{N} \delta_{a_j,c} \delta_{b_j,d}$, which represents the number of atoms populating the local density-matrix element $|c\rangle\langle d|$ in a full density matrix element $|\alpha\rangle \langle \beta|$ written in the product basis ($a_j$ and $b_j$ are distinct elements of $\alpha$ and $\beta$, respectively). Conservation of the total atom number then leads to $\sum_{c,d} n_{cd}=N$. In this representation, we can use a symmetric operator $\hat{\mathbf{n}}$ to denote the permutation-invariant mixed states.
	
    Comparing with the original dimension of the density matrix $\mathcal{O}(9^{N})$, the computational complexity is reduced to $C_{N+8}^{8}=(N+8)!/(8!N!)=\mathcal{O}(N^8)$ in this  representation, allowing us to diagonalize the Liouvillian at a much larger system size. Based on this representation, we can also efficiently solve the dynamics and obtain the population of the state $|\alpha\rangle$ with the matrix elements $\hat{\rho}_{\alpha,\alpha}$. Based on the mapping, we can denote $\hat{\rho}_{(n_{00},0,0,n_{++},0,n_{--},0,0,0)}$ as $P_{(n_{00},n_{++},n_{--})}$, which indicates $n_{00}$, $n_{++}$ and $n_{--}$ atoms in $|0\rangle$, $|+\rangle$ and $|-\rangle$, respectively. Populations $\mathcal{P}_\mu$ ($\mu=0,+,-$) of the state $|0\rangle$, $|+\rangle$ and $|-\rangle$ are then given by
	\begin{align}
		\mathcal{P}_0&=\sum_{n_{00}=1}^{N} \sum_{n_{++}=0}^{N-n_{00}} C_{N-1}^{n_{00}-1} C_{N-n_{00}}^{n_{++}} P_{(n_{00},n_{++},N-n_{00}-n_{++})}, \nonumber\\
		\mathcal{P}_+&=\sum_{n_{++}=1}^{N} \sum_{n_{00}=0}^{N-n_{++}} C_{N-1}^{n_{++}-1} C_{N-n_{++}}^{n_{00}} P_{(n_{00},n_{++},N-n_{00}-n_{++})}, \nonumber\\
		\mathcal{P}_-&=\sum_{n_{--}=1}^{N} \sum_{n_{--}=0}^{N-n_{--}} C_{N-1}^{n_{--}-1} C_{N-n_{--}}^{n_{++}} P_{(N-n_{++}-n_{--},n_{++},n_{--})}.
	\end{align}
In the main text, we utilize this approach to obtaining the low-lying Liouvillian spectra and the two-time correlation fuction up to $N=20$.

	\section{Cumulant expansion}\label{AppC}
For a generic power-law interaction, the system is not permutation invariant, so we instead use cumulant expansions to study the dynamics. The second-order cumulant expansions used in the main text go beyond the mean-field theory by assuming Gaussian-type quantum correlations, i.e.,
	\begin{align}
		\label{cumulant1}
		\langle\hat{\sigma}_{a}^{\alpha}\hat{\sigma}_{b}^{\beta}\hat{\sigma}_{c}^{\gamma}\rangle &\approx \langle\hat{\sigma}_{a}^{\alpha}\hat{\sigma}_{b}^{\beta}\rangle\langle\hat{\sigma}_{c}^{\gamma}\rangle + \langle\hat{\sigma}_{a}^{\alpha}\hat{\sigma}_{c}^{\gamma}\rangle\langle\hat{\sigma}_{b}^{\beta}\rangle + \langle\hat{\sigma}_{b}^{\beta}\hat{\sigma}_{c}^{\gamma}\rangle\langle\hat{\sigma}_{a}^{\alpha}\rangle  -2\langle\hat{\sigma}_{a}^{\alpha}\rangle\langle\hat{\sigma}_{b}^{\beta}\rangle\langle\hat{\sigma}_{c}^{\gamma}\rangle .
	\end{align}
	With the above truncation of the third-order correlations, we can obtain a closed set of equations for the first and the second moments. The typical terms are given by 
	\begin{align}
		\partial_t\langle\hat{\sigma}_{a}^{0+}\rangle =&\  i\frac{\Omega}{2}[\langle\hat{n}_{a}^{+}\rangle-\langle\hat{n}_{a}^{0}\rangle+\langle\hat{\sigma}_{a}^{+-}\rangle^{\ast}]+i\left(\Delta_+ +i\frac{\gamma}{2}\right)\langle\hat{\sigma}_{a}^{0+}\rangle  -i\sum_{j\neq a}V_{aj}[\langle\hat{\sigma}_{a}^{0+}\rangle-\langle\hat{n}_{j}^{0}\hat{\sigma}_{a}^{0+}\rangle
		], \\	
		\partial_t\langle\hat{\sigma}_{a}^{0-}\rangle =&\ i\frac{\Omega}{2}[1-\langle\hat{n}_{a}^{+}\rangle-2\langle\hat{n}_{a}^{0}\rangle+\langle\hat{\sigma}_{a}^{+-}\rangle]+i\left(\Delta_- +i\frac{\gamma}{2}\right)\langle\hat{\sigma}_{a}^{0-}\rangle -i\sum_{j\neq a}V_{aj}[\langle\hat{\sigma}_{a}^{0-}\rangle-\langle\hat{n}_{j}^{0}\hat{\sigma}_{a}^{0-}\rangle
		],     \\
		\partial_t\langle\hat{n}_{a}^{0}\hat{\sigma}_{b}^{0+}\rangle =&\
		i\frac{\Omega}{2}[-\langle\hat{\sigma}_{a}^{0+}\hat{\sigma}_{b}^{0+}\rangle-\langle\hat{\sigma}_{b}^{0+}\hat{\sigma}_{a}^{0-}\rangle+\langle\hat{\sigma}_{b}^{0+}\hat{\sigma}_{a}^{+0}\rangle+\langle\hat{\sigma}_{b}^{0+}\hat{\sigma}_{a}^{-0}\rangle-\langle\hat{n}_{a}^{0}\hat{n}_{b}^{0}\rangle+\langle\hat{n}_{a}^{0}\hat{n}_{b}^{+}\rangle+\langle\hat{n}_{a}^{0}\hat{\sigma}_{b}^{+-}\rangle^{\ast}] \nonumber\\
		&+i\left(\Delta_+ +i\frac{\gamma}{2}\right)\langle\hat{n}_{a}^{0}\hat{\sigma}_{b}^{0+}\rangle+\gamma[\langle\hat{\sigma}_{b}^{0+}\rangle-\langle\hat{n}_{a}^{0}\hat{\sigma}_{b}^{0+}\rangle] \nonumber\\
		&-i\sum_{j\neq a,b}V_{bj}[\langle\hat{n}_{a}^{0}\hat{\sigma}_{b}^{0+}\rangle-\langle\hat{n}_{a}^{0}\hat{\sigma}_{b}^{0+}\rangle\langle\hat{n}_{j}^{0}\rangle-\langle\hat{n}_{a}^{0}\hat{n}_{j}^{0}\rangle\langle\hat{\sigma}_{b}^{0+}\rangle-\langle\hat{n}_{j}^{0}\hat{\sigma}_{b}^{0+}\rangle\langle\hat{n}_{a}^{0}\rangle+2\langle\hat{n}_{a}^{0}\rangle\langle\hat{\sigma}_{b}^{0+}\rangle\langle\hat{n}_{j}^{0}\rangle
		],  \label{S14} \\
		\partial_t\langle\hat{n}_{a}^{0}\hat{\sigma}_{b}^{0-}\rangle =&\
		i\frac{\Omega}{2}[-\langle\hat{\sigma}_{a}^{0+}\hat{\sigma}_{b}^{0-}\rangle-\langle\hat{\sigma}_{a}^{0-}\hat{\sigma}_{b}^{0-}\rangle+\langle\hat{\sigma}_{a}^{0+}\hat{\sigma}_{b}^{-0}\rangle^{\ast}+\langle\hat{\sigma}_{b}^{0-}\hat{\sigma}_{a}^{-0}\rangle-2\langle\hat{n}_{a}^{0}\hat{n}_{b}^{0}\rangle-\langle\hat{n}_{a}^{0}\hat{n}_{b}^{+}\rangle+\langle\hat{n}_{a}^{0}\rangle+\langle\hat{n}_{a}^{0}\hat{\sigma}_{b}^{+-}\rangle] \nonumber\\
		&+i\left(\Delta_- +i\frac{\gamma}{2}\right)\langle\hat{n}_{a}^{0}\hat{\sigma}_{b}^{0-}\rangle+\gamma[\langle\hat{\sigma}_{b}^{0-}\rangle-\langle\hat{n}_{a}^{0}\hat{\sigma}_{b}^{0-}\rangle] \nonumber\\
		&-i\sum_{j\neq a,b}V_{bj}[\langle\hat{n}_{a}^{0}\hat{\sigma}_{b}^{0-}\rangle-\langle\hat{n}_{a}^{0}\hat{\sigma}_{b}^{0-}\rangle\langle\hat{n}_{j}^{0}\rangle-\langle\hat{n}_{a}^{0}\hat{n}_{j}^{0}\rangle\langle\hat{\sigma}_{b}^{0-}\rangle-\langle\hat{n}_{j}^{0}\hat{\sigma}_{b}^{0-}\rangle\langle\hat{n}_{a}^{0}\rangle+2\langle\hat{n}_{a}^{0}\rangle\langle\hat{\sigma}_{b}^{0-}\rangle\langle\hat{n}_{j}^{0}\rangle
		].
	\end{align}
The remaining equations can be written in a similar manner. Eventually, we can arrive at a closed set of equations containing $5N$ first moments and $21N(N-1)$ second moments. In this way, the complexity of the many-body problem reduces from $\mathcal{O}(9^N)$ to $\mathcal{O}(N^2)$ \cite{ddTC4}. For an all-to-all interaction, the problem is, however, independent of the system size $N$, which appears as a parameter in a set of 26 equations.
	
	To further reduce the complexity for distance-dependent interactions, we consider periodic boundary conditions and apply translational invariance, which reduces the computational complexity from $\mathcal{O}(N^2)$ to $\mathcal{O}(N)$. Specifically, we assume that the first moments have a homogeneous distribution, and the second moments only depend on the distance between the two sites $a$ and $b$. Taking Eq.~\eqref{S14} as an example, the equation is reduced to a form that only depends on $n=a-b$, i.e.,
    \begin{align}
    \partial_t\langle\hat{n}^{0}\hat{\sigma}^{0+}\rangle_{n} =&\
		i\frac{\Omega}{2}[-\langle\hat{\sigma}^{0+}\hat{\sigma}^{0+}\rangle_{n}-\langle\hat{\sigma}^{0+}\hat{\sigma}^{0-}\rangle_{-n}+\langle\hat{\sigma}^{0+}\hat{\sigma}^{+0}\rangle_{-n}+\langle\hat{\sigma}^{0+}\hat{\sigma}^{-0}\rangle_{-n}-\langle\hat{n}^{0}\hat{n}^{0}\rangle_{n}+\langle\hat{n}^{0}\hat{n}^{+}\rangle_{n}+\langle\hat{n}^{0}\hat{\sigma}^{+-}\rangle_{n}^{\ast}] \nonumber\\
		&+i\left(\Delta_++i\frac{\gamma}{2}\right)\langle\hat{n}^{0}\hat{\sigma}^{0+}\rangle_{n}+\gamma[\langle\hat{\sigma}^{0+}\rangle-\langle\hat{n}^{0}\hat{\sigma}^{0+}\rangle_{n}] \nonumber\\
		&-i\sum_{\mu\neq -n,0}V_{\mu}[\langle\hat{n}^{0}\hat{\sigma}^{0+}\rangle_{n}-\langle\hat{n}^{0}\hat{\sigma}^{0+}\rangle_{n}\langle\hat{n}^{0}\rangle-\langle\hat{n}^{0}\hat{n}^{0}\rangle_{n+\mu}\langle\hat{\sigma}^{0+}\rangle-\langle\hat{n}^{0}\hat{\sigma}^{0+}\rangle_{-\mu}\langle\hat{n}^{0}\rangle+2\langle\hat{n}^{0}\rangle\langle\hat{\sigma}^{0+}\rangle\langle\hat{n}^{0}\rangle
		]. 
    \end{align}
    Such a treatment allows us to simulate the dynamical evolution of the system containing thousands of atoms. 
    
\begin{figure}
		\centering
		\includegraphics[width=0.8\linewidth]{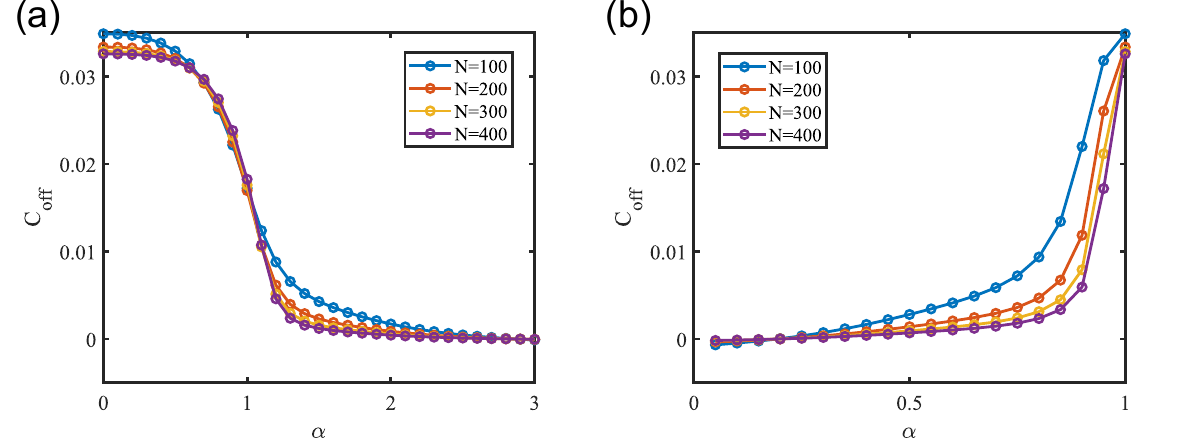}
		\caption{The diagnostics in different distance-dependent long-range interactions. Different color curves represent different system size $N$ from 100 to 400. (a) and (b) are the long-range off-diagonal correlation in the steady state $\mathcal{C}_\mathrm{off}$ versus parameter $\alpha$ in power-law decaying and exponentially decaying interactions, respectively.} 
		\label{fig:Fig.S2}
	\end{figure}

    As mentioned in the main text, for a sufficiently long-ranged interaction, the BTC phase is preserved, which can be revealed by the long-range off-diagonal correlation of the steady state
	\begin{equation}
		\mathcal{C}_\mathrm{off}=\frac{1}{N(N-1)}\sum_{i\neq j}\left[\langle s_i^z s_j^z\rangle_\mathrm{ss}-\langle s_i^z\rangle_\mathrm{ss} \langle s_j^z\rangle_\mathrm{ss} \right].
	\end{equation}
To demonstrate that a long-range interaction is indeed important to stabilize the BTC, we compare the behaviors of a power-law decaying interaction $V_{ij} = C/|i-j|^\alpha$ with an exponentially decaying interaction $V_{ij} = C/\alpha^{|i-j|}$. As shown in Fig.~\ref{fig:Fig.S2}(a), for a power-law interaction, as the system size $N$ increases, neighboring curves intersect and the crossing point appears to converge, indicating a finite critical value $\alpha_c$ in the thermodynamic limit. However, for an exponentially decaying interaction [see Fig.~\ref{fig:Fig.S2}(b)], as $N$ increases, different curves do not intersect but continue approaching zero for any $\alpha<1$, indicating the absence of a BTC phase.
	
	\section{Quantum-trajectory-based matrix product state}\label{AppD}

 To verify the validity of the cumulant expansions and confirm its qualitative correctness, we further implement quantum-trajectory-based matrix product state (MPS) simulations. MPS is an efficient representation of many-body states with area-law entanglement. The combination of MPS and quantum trajectory approaches offers two key advantages here: (1) the quantum jump can suppress the rapid growth of the MPS bond dimensions during the time evolution; and (2) different quantum trajectory samplings can be accelerated through parallelization.

\begin{figure}
		\centering
		\includegraphics[width=\linewidth]{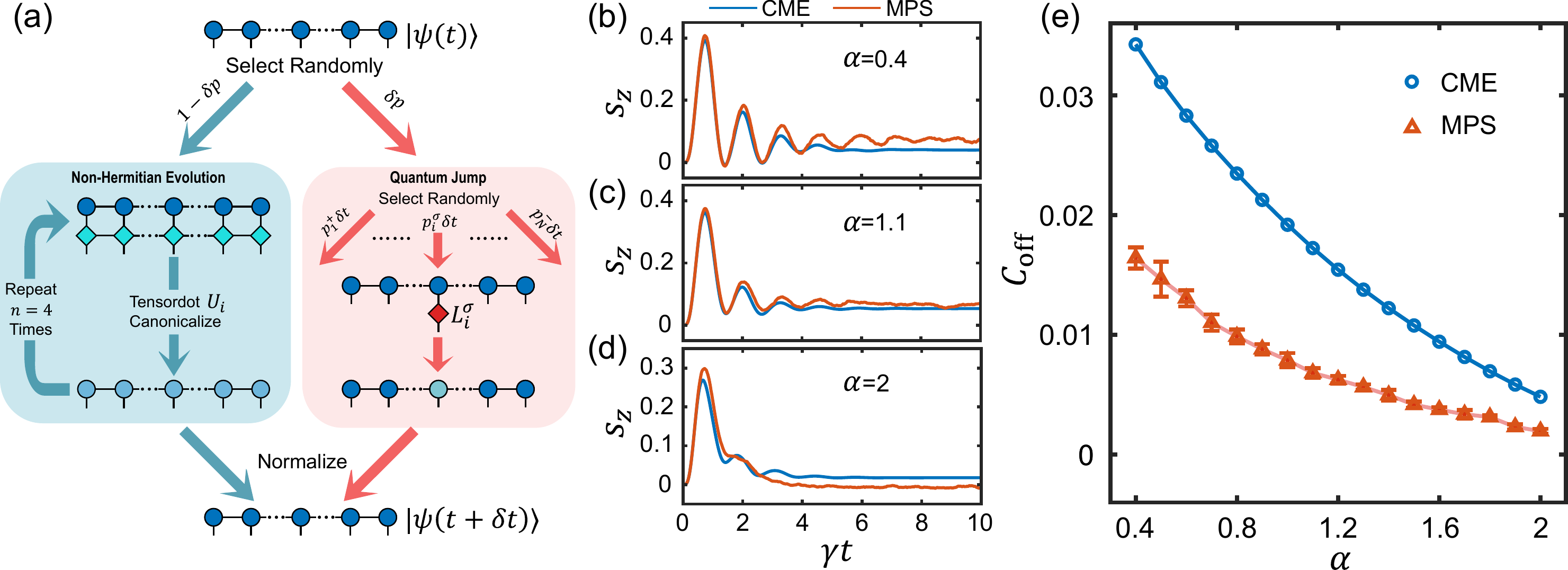}
		\caption{(a) Illustration of the single-step time evolution of quantum-trajectory-based matrix product state. (b), (c) and (d) show the time evolution of the order parameter $s_z$ with different $\alpha$ calculated by CME and MPS. (e) shows the relation between steady-state off-diagonal correlation $C_{\text{off}}$ and $\alpha$. The errorbars indicate the temporal fluctuations caused by finite trajectory realizations.}
		\label{fig:MPS}
	\end{figure}

Our quantum trajectory simulations employ the quantum jump approach. The total evolution time $T$ is discretized into $n$ intervals with a duration $\delta t = T/n$. As illustrated in Fig.~\ref{fig:MPS}(a), the system evolution follows:
\begin{equation}
\partial_t\rho = -i[\mathcal{H}_{\text{eff}}, \rho] + \sum_{j=1}^{N}\sum_{\sigma=\pm} L_j^{\sigma}\rho L_j^{\sigma\dagger},
\label{eq:master}
\end{equation}
where the effective Hamiltonian is $\mathcal{H}_{\text{eff}} = H - \frac{i}{2}\sum_j \sum_{\sigma=\pm}L_j^{\sigma\dagger} L_j^{\sigma}$. For a pure initial state $\ket{\psi(t)}$, the time evolution of the system can be separated into two parts. For the non-Hermitian evolution, it reads
\begin{equation}
|\phi(t+\delta t)\rangle = (1-i\mathcal{H}_{\text{eff}}\delta t)|\psi(t)\rangle.
\label{eq:no_jump}
\end{equation}
The non-Hermiticity leads to a non-conservation of the inner product, i.e., $\langle \phi(t+\delta t)|\phi(t+\delta t)\rangle=1-\delta p$, where $\delta p=\delta t\sum_j \sum_{\sigma=\pm}p_{j}^{\sigma},p_{j}^{\sigma}=\langle\psi(t)|L_j^{\sigma\dagger} L_j^{\sigma}|\psi(t)\rangle$. Including the quantum jump term, the discrete time evolution of the density matrix can be rewritten as
\begin{equation}
\rho(t+\delta t)=(1-\delta p)|\phi'(t+\delta t)\rangle\langle \phi'(t+\delta t)|+\delta p\sum_j \sum_{\sigma=\pm}\frac{p_{j}^{\sigma}}{\sum_{j'} \sum_{\sigma'=\pm}p_{j'}^{\sigma'}}|\psi_{j}^{\sigma}(t)\rangle\langle \psi_{j}^{\sigma}(t)|,
\label{eq:ev}
\end{equation}
where $|\phi'(t+\delta t)\rangle=|\phi(t+\delta t)\rangle/\sqrt{1-\delta p}$ and $|\psi_{j}^{\sigma}(t)\rangle=L_j^{\sigma}|\psi(t)\rangle/\sqrt{p_j^{\sigma}}$ are normalized quantum states. This demonstrates that after an evolution time $\delta t$, the density matrix of the system becomes a statistical mixture of $|\phi'(t+\delta t)\rangle$ and $|\psi_{j}^{\sigma}(t)\rangle$: The initial state evolves either under the effective Hamiltonian $\mathcal{H}_{\text{eff}}$ with a probability $1-\delta p$, or undergoes a quantum jump with a probability $\delta p$, collapsing the state into $|\psi_{j}^{\sigma}(t)\rangle$ with conditional probability $p_{j}^{\sigma}$. 

Repeating this procedure as shown in Fig.~\ref{fig:MPS}(a), we obtain a trajectory $|\psi_i(t)\rangle$ that evolves from $0$ to $T$. Each trajectory $\{|\psi_i(t)\rangle,i=1,2,3\cdots\}$ is independent and can be executed in parallel. The observable is then averaged over different trajectory realizations $\langle \hat o(t)\rangle=\sum_i\langle \psi_i(t)|\hat o|\psi_i(t)\rangle/N_{\text{tr}}$, where $N_{\text{tr}}$ is the number of trajectories.

In the MPS implementation, the quantum jump can be realized by simply applying the local jump operator $L_j^{\sigma}$ to the state. The time evolution governed by the effective Hamiltonian is more involved. In our system, the effective Hamiltonian itself can be expressed as a matrix product operator (MPO) in the form
\begin{equation}
\mathcal{M}=\begin{pmatrix}
    \hat I&\hat{\textbf{C}}&\hat H_{\text{eff,local}}\\\textbf{0}&\hat {\textbf{A}}&\hat{\textbf{B}}\\0&\textbf{0}&\hat I
    \end{pmatrix}.
\end{equation}
Here, the blocks $\hat {\textbf{A}},\hat {\textbf{B}},\hat {\textbf{C}}$ depend on the types of the long-range interaction, while
\begin{equation}
\mathcal{\hat H}_{\text{eff, local}}=\frac{\Omega}{\sqrt{2}} \hat s_x-\Delta \hat s_z^2-E\hat s_z-\frac{i\gamma}{2} \hat s_z^2
\end{equation}
denotes the local component of the effective Hamiltonian. Following Ref.~\cite{mpsevolution}, the single-step time evolution can be approximated up to $n$ orders, $\psi(t+\delta t)=\mathcal{U}\psi(t)=\mathcal{U}_n\mathcal{U}_{n-1}\cdots\mathcal{U}_{1}\psi(t)$, where $\mathcal{U}$ is the time evolving operator factorized into products of $\mathcal{U}_n$, whose MPO form is given by
\begin{equation}
    \mathcal{U}_n=\begin{pmatrix}
\hat I-i z_n\delta t\mathcal{\hat H}_{\text{eff, local}}&\sqrt{-iz_n\delta t}\hat{\textbf{C}} \\
\sqrt{-iz_n\delta t}\hat{\textbf{B}}&\hat{\textbf{A}}
\end{pmatrix},
\end{equation}
where the complex coefficient $z_n$ should satisfy $\sum_{n}z_n=1,\sum_{n<m}z_n z_m=1/2,\sum_{n}z^2_n=0$. In this work, we adopt a fourth-order ($n=4$) approximation. 

For the simplest all-to-all interaction, $\hat {\textbf{A}}=\hat I,~\hat {\textbf{B}}=\hat n=\hat s_z^2,~\hat {\textbf{C}}=V\hat n$ with $V=\chi/(N-1)$. For the simulation in Fig.~4(a) of the main text, the bond dimension $D=30$, and the number of trajectories are $N_{\text{tr}}=500,200,100,50$ for $N=20,40,60,80$. In the case of a power-law interaction $V(r)=V_0r^\alpha$, the MPO has a much larger dimension. Here, the power-law interaction is approximated as a superposition of exponentially decaying interactions\cite{mpsevolution}, i.e.,
\begin{equation}
    \label{fit-exp}
    V(r)\approx\sum_{k=1}^{m}a_kV_k^{r-1}+c.
\end{equation}
The coefficients $a_k, V_k, c$ are fitted to yield a smallest deviation from the power-law interaction. The accuracy of the fitting depends on the number of terms $m$, which is related to both the system size $N$ and the power-law exponent $\alpha$. The resulting matrix blocks are given by
\begin{equation}
    ~\hat {\textbf{A}}=\begin{pmatrix}
        V_1\hat I&0&\dots&0&0\\
        0&V_2\hat I&\dots&0&0\\
        \vdots&\vdots&\ddots&\vdots&\vdots&\\
        0&0&\dots&V_m\hat I&0\\
        0&0&\dots&0&\hat I\\
    \end{pmatrix},~\hat{{\textbf{B}}}=\begin{pmatrix}
    \hat n\\\hat n\\\vdots\\\hat n\\\hat n
    \end{pmatrix},~\hat{{\textbf{C}}}=\begin{pmatrix}
        a_1\hat {n}&a_2\hat {n}&\dots&a_m \hat {n}&c\hat {n}
    \end{pmatrix}.
\end{equation}

Figures~\ref{fig:MPS}(b)-(e) show the results for a power-law interaction with $N=20$ and $N_{\text{tr}}=200$. For the order parameter $s_z=\sum_i^N s_i^z/N$, we compare the time evolution calculated by CME and MPS [see Figs.~\ref{fig:MPS}(b)-(d)], and find that the CME can always yield a nice agreement with the MPS for the entire range of $\alpha$ considered in the main text. For the quantum correlation $C_{\text{off}}$ [see Fig.~\ref{fig:MPS}(e)], the non-Gaussianity of the steady state yields a visible discrepancy between the CME and the MPS. However, similar to the case of all-to-all interactions, the trend of $\mathcal{C}_\mathrm{off}$ as a function of $\alpha$ is well captured by the CME. This qualitative agreement enhances the credibility of the CME-based analysis presented in the main text.
    
\bibliography{ref}